\documentclass[conference]{IEEEtran}
\IEEEoverridecommandlockouts

\usepackage{cite}
\usepackage{amsmath,amssymb,amsfonts}
\usepackage{algorithmic}
\usepackage{graphicx}
\usepackage{textcomp}
\usepackage{xcolor}
\usepackage{booktabs}
\usepackage{siunitx}
\usepackage{multirow}
\usepackage{bm}
\usepackage[T1]{fontenc}
\usepackage{multirow, array}
\usepackage{amsmath, amssymb}
\usepackage{siunitx}
\usepackage{subcaption}
\usepackage{csvsimple}
\usepackage{graphicx}
\graphicspath{{assets/figures/}{assets/logos/}}
\usepackage{booktabs, xcolor, makecell}
\usepackage{tabularx}
\newcolumntype{Y}{>{\raggedright\arraybackslash}X}
\usepackage[normalem]{ulem}
\usepackage{float}

\def\BibTeX{{\rm B\kern-.05em{\sc i\kern-.025em b}\kern-.08em
    T\kern-.1667em\lower.7ex\hbox{E}\kern-.125emX}}
\begin{document}

\title{Investigation on the Robustness of Acoustic
Foundation Models on Post Exercise Speech
\thanks{This work was supported in part by the NVIDIA Academic Award Grant.}
}

\author{\IEEEauthorblockN{Xiangyuan Xue$^{\dag, \ddag}$, Yuyu Wang$^{\dag}$, Ruijie Yao$^*$, Xiaoyue Ni$^*$, Xiaofan Jiang$^{**}$, and Jingping Nie$^{\dag}$}
\IEEEauthorblockA{
\textit{$^{\dag}$University of North Carolina at Chapel Hill, Chapel Hill, NC, USA; $^{\ddag}$ University of Auckland, Auckland, NZ}\\
\textit{$^*$Duke University, Durham, NC, USA; $^{**}$Columbia University, New York, NY, USA}\\
{\{xyxue, yuyuwang, jingping\}@unc.edu}, \{ruijie.yao, xiaoyue.ni\}@duke.edu, xiaofan.jiang@ee.columbia.edu}


}

\maketitle

\begin{abstract}
Automatic speech recognition (ASR) has been extensively studied on neutral and stationary speech, yet its robustness under post-exercise physiological shift remains underexplored. Compared with resting speech, post-exercise speech often contains micro-breaths, non-semantic pauses, unstable phonation, and repetitions caused by reduced breath support, making transcription more difficult. In this work, we benchmark acoustic foundation models on post-exercise speech under a unified evaluation protocol. We compare sequence-to-sequence models (Whisper and FunASR/Paraformer) and self-supervised encoders with CTC decoding (Wav2Vec2, HuBERT, and WavLM), under both off-the-shelf inference and post-exercise in-domain fine-tuning. Across the \texttt{Static}/\texttt{Post-All} benchmark, most models degrade on post-exercise speech, while FunASR shows the strongest baseline robustness at 14.57\% WER and 8.21\% CER on \texttt{Post-All}. Fine-tuning substantially improves several CTC-based models, whereas Whisper shows unstable adaptation. As an exploratory case study, we further stratify results by fluent and non-fluent speakers; although the non-fluent subset is small, it is consistently more challenging than the fluent subset. Overall, our findings show that post-exercise ASR robustness is strongly model-dependent, that in-domain adaptation can be highly effective but not uniformly stable, and that future post-exercise ASR studies should explicitly separate fluency-related effects from exercise-induced speech variation.

\end{abstract}


\section{introduction}~\label{sec:intro}
Speech is a practical sensing modality for real-world health and activity monitoring because it can be captured with widely available devices such as smartphones, laptops, earbuds, and wearables~\cite{xia2025convergence}. Unlike many biosignals, which often require specialized hardware or body contact, speech can be collected at scale with minimal user burden. These advantages have motivated growing interest in speech-based assessment for respiratory monitoring, stress and fatigue estimation, symptom tracking, and exercise-related coaching. Recent studies further suggest that post-exercise speech contains physiologically meaningful information, including cues related to breathing patterns, exertion level, and cardiorespiratory recovery, making it a promising signal for fitness monitoring, rehabilitation, and telehealth applications \cite{nie2025multimodal,mitra2024breathing,zhou2025voice, wang2025breathing, nie2025soundtrack}.

\begin{figure}[t!]
    \centering
    \includegraphics[width=\linewidth]{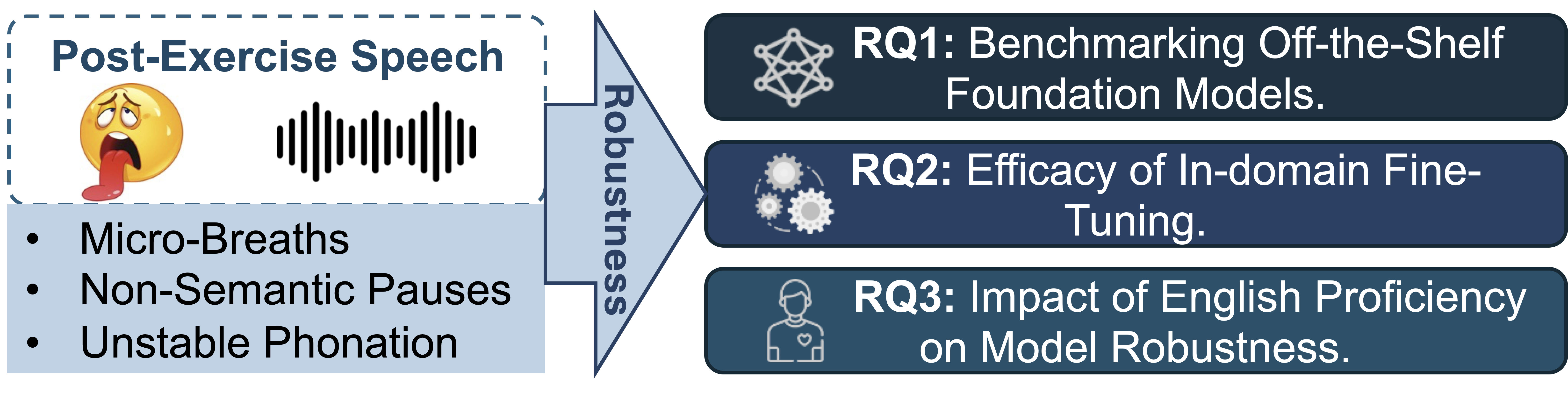}
    \caption{Framework for evaluating and enhancing ASR robustness on post-exercise speech.}
    \label{fig:teasure}
    \vspace{-0.2in}
\end{figure}

Despite its promise for health sensing, post-exercise speech poses significant challenges for automatic speech recognition (ASR). Compared with resting speech, post-exercise utterances often contain micro-breaths, irregular pauses, reduced breath support, unstable phonation, repetitions, hesitations, and truncated words, introducing variability across acoustic, temporal, and lexical levels~\cite{nie2025multimodal, wang2025breathing, mitra2025leveraging}. Heavy breathing and unstable voicing may distort patterns learned during pretraining, while irregular pauses and speaking-rate variation can disrupt alignment between acoustic frames and textual output. Repetitions and incomplete productions further increase the mismatch between intended and realized speech. Consequently, post-exercise speech represents a realistic yet underexplored source of distribution shift for modern ASR systems. This issue is practically important because ASR often serves as the front end for downstream health-related inference, such as symptom summarization, recovery-state analysis, and digital coaching~\cite{nie2025foundation, mitra2024investigating}. More broadly, robustness under physiological speech variation is also relevant to downstream audio-language systems for speech understanding and caption generation \cite{xue2026edge}.In many real-world scenarios, including exercise recovery, rehabilitation sessions, and remote health check-ins, users may interact with voice systems while short of breath, making robust recognition under physiological stress essential~\cite{milne2020effectiveness}. However, most ASR benchmarks remain centered on read or conversational speech rather than speech produced under physiological perturbation, leaving the robustness of current models to post-exercise speech insufficiently characterized. In this setting, two questions arise: how much ASR performance degrades from static to post-exercise speech under off-the-shelf inference, and whether post-exercise fine-tuning can recover this loss.

One related line of research examines ASR performance on atypical or dysfluent speech, particularly stuttered speech~\cite{lea2021sep28k,ratner2018fluencybank,mujtaba2024lost,lea2023enabling,mujtaba2024inclusive,mujtaba2025finetuning}. These studies show that ASR systems often degrade when speech departs from the fluent, well-articulated conditions emphasized in conventional benchmarks, although targeted adaptation can partially mitigate the issue. This literature is relevant because post-exercise speech may also contain repetitions, hesitations, and timing irregularities. However, the underlying mechanism differs: stuttered speech reflects a persistent speech production disorder, whereas post-exercise speech arises from a transient physiological state induced by exertion. Prior work, therefore, highlights ASR's vulnerability to atypical speech but does not directly address how current models behave during exercise-induced state shifts.

To address this gap and better understand the robustness of modern ASR systems under exercise-induced speech variation, we conduct a systematic study of acoustic foundation models on post-exercise speech. We compare two major architectural families used in contemporary ASR. The first consists of sequence-to-sequence models, represented by Whisper~\cite{radford2022whisper} and FunASR/Paraformer~\cite{gao2022paraformer}, which directly decode acoustic input into text. The second consists of self-supervised acoustic encoders with CTC decoding, represented by Wav2Vec2~\cite{baevski2020wav2vec2} and WavLM \cite{chen2021wavlm}. These families differ not only in architecture but also in how they handle timing variation, alignment, and domain mismatch, making them well-suited for studying domain adaptation under physiological speech shifts. To examine both architectural and scale effects, we evaluate small and large variants where available, resulting in a unified nine-model benchmark. 

Our exploratory study is organized around three questions, shown in Figure~\ref{fig:teasure}: (\emph{i}) \textbf{RQ1:} How much does ASR performance degrade from static to post-exercise speech under off-the-shelf inference? (\emph{ii}) \textbf{RQ2:} Can post-exercise fine-tuning recover this degradation, and does the outcome depend on model architecture? (\emph{iii}) \textbf{RQ3:} Does the pattern of robustness and adaptation differ between fluent and non-fluent English speakers? This analysis helps characterize how post-exercise ASR robustness varies across speakers.

The results in our exploratory study show that post-exercise robustness does not transfer uniformly across ASR models. Under off-the-shelf inference, most systems perform worse on post-exercise speech, but the magnitude of degradation varies substantially across architectures. For example, Wav2Vec2-Base increases from 21.53\% to 32.30\% WER and from 10.11\% to 15.70\% CER, HuBERT-Large increases from 11.54\% to 19.90\% WER and from 5.49\% to 9.52\% CER, and WavLM-Base+ increases from 35.32\% to 45.45\% WER and from 14.78\% to 20.46\% CER. By contrast, FunASR remains comparatively stable, changing only from 12.29\% to 14.57\% WER and from 6.70\% to 8.21\% CER, and achieves the strongest post-exercise baseline result. Post-exercise fine-tuning is likewise not uniformly beneficial.It substantially improves several self-supervised learning (SSL) models with connectionist temporal classification (CTC) decoding, reducing WER/CER from 45.45\%/20.46\% to 22.94\%/8.67\% for WavLM-Base+ and from 45.98\%/21.04\% to 24.75\%/9.33\% for HuBERT-Base, but remains unstable for some sequence-to-sequence settings, most notably Whisper-Base, which degrades from 27.48\% to 58.11\% WER and from 21.87\% to 47.62\% CER after fine-tuning. We further observe that post-exercise recognition is consistently more difficult for \texttt{Non-Fluent} speakers than for \texttt{Fluent} speakers. These findings suggest that robustness under physiological speech variation depends not only on model scale or pretraining strength, but also on architectural properties that shape alignment, timing sensitivity, and adaptation behavior, with implications for both domain adaptation and human-centered speech processing. The three contributions in this study are summarized below:
\begin{itemize}
    \item We provide the first \emph{unified benchmark} of acoustic foundation models under post-exercise physiological shift, showing that ASR performance consistently degrades from static to post-exercise speech, with substantial variability across architectures and model scales.
    \item We show that \emph{post-exercise fine-tuning} can effectively recover performance for several SSL+CTC models, but the benefit is highly architecture-dependent and not uniformly stable, with some sequence-to-sequence models (e.g., Whisper) showing degradation after adaptation.
    \item We demonstrate an exploratory \emph{fluency-based case study}, showing that non-fluent speakers are consistently more challenging under both static and post-exercise conditions, and that exercise-induced variation interacts with pre-existing fluency-related difficulty, motivating the need for subgroup-aware evaluation in future ASR studies.
\end{itemize}

\section{Datasets and Data Preparation}~\label{sec:data}
\begin{figure}[t!]
    \centering
    \includegraphics[width=\columnwidth]{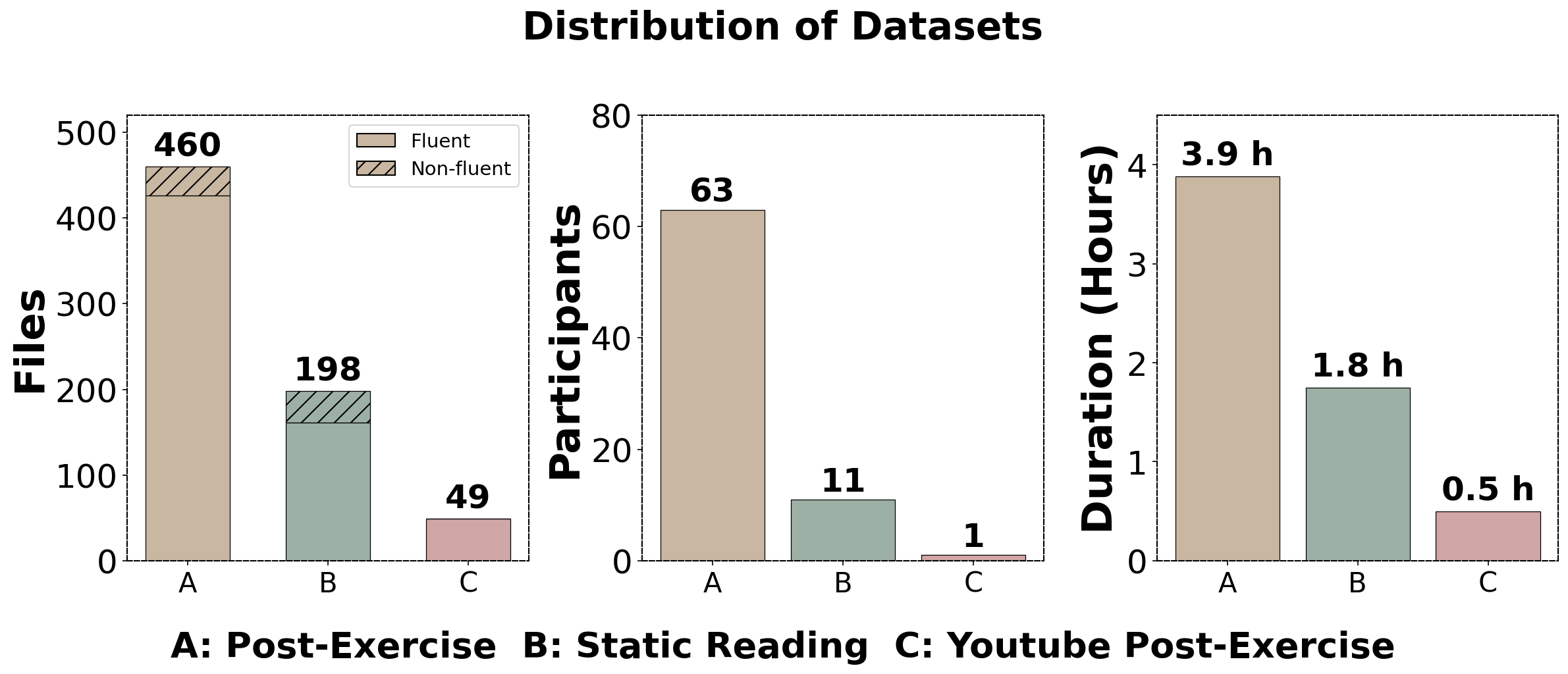}
    \caption{Dataset distribution.}
    \label{fig:dataset_distribution}

\end{figure}

We evaluate ASR robustness under post-exercise physiological shift using three complementary datasets spanning both controlled and in-the-wild recording conditions: (\textbf{A}) \texttt{Post-Exercise}, comprising 388 structured and unstructured recordings (approximately 3.3 hours) from 59 participants, including both fluent and non-fluent speakers, collected after cardio exercise and published in \cite{nie2025multimodal}, together with an additional 72 structured recordings (approximately 0.6 hours) from 4 fluent speakers; (\textbf{B}) \texttt{Static Reading}, consisting of 198 structured paragraph-reading recordings (approximately 1.8 hours) from 11 fluent and non-fluent speakers; and (\textbf{C}) \texttt{YouTube Post-Exercise}, consisting of 49 unstructured free-speech recordings (approximately 0.5 hours) from a single speaker recorded in an in-the-wild setting. The detailed distribution of the three datasets is illustrated in Figure~\ref{fig:dataset_distribution}. Apart from the publicly available and online data, all remaining recordings were collected locally under an approved Institutional Review Board (IRB) protocol. Collectively, these datasets span multiple axes of variation, including speaking task (structured/unstructured), exercise condition (post-exercise/static), speaker population (fluent/non-fluent), and recording context, enabling us to characterize post-exercise ASR as a robustness problem under diverse real-world and semi-controlled conditions. 

Considering the dataset sizes, we merge (\textbf{A}) \texttt{Post-Exercise} and (\textbf{C}) \texttt{YouTube Post-Exercise} into a unified set, denoted as \texttt{Post-All}, while (\textbf{B}) \texttt{Static Reading} is used as the reference set, denoted as \texttt{Static}. The \texttt{Static} set contains 198 recordings from 11 speakers, while \texttt{Post-All} contains 509 recordings from 64 speakers. Using these two datasets, we address the three research questions (\textbf{RQ1}–\textbf{RQ3}) outlined in Figure~\ref{fig:teasure} and Section~\ref{sec:intro}. We conduct a comprehensive evaluation across nine commonly used ASR models, as described in Section~\ref{sec:experiment} and summarized in Table~\ref{tab:model_specs}.

\section{Experimental Setup}~\label{sec:experiment}
\noindent\textbf{ASR Models. }
As summarized in Table~\ref{tab:model_specs}, we benchmark 9 ASR models spanning 5 model families across 2 architectural families: (I) sequence-to-sequence (seq2seq) encoder--decoder models, represented by Whisper \cite{radford2022whisper} and FunASR/Paraformer \cite{gao2022paraformer}, which map acoustic input directly to text tokens within a unified end-to-end decoding framework and often provide strong off-the-shelf performance due to large-scale pretraining and generative decoding; and (II) self-supervised learning (SSL) acoustic encoders with connectionist temporal classification (CTC), represented by Wav2Vec2 \cite{baevski2020wav2vec2}, HuBERT \cite{hsu2021hubert}, and WavLM \cite{chen2021wavlm}, where encoders are first pretrained on large-scale unlabeled speech to learn general acoustic representations and then adapted to transcription through supervised fine-tuning. This distinction is especially relevant for post-exercise speech, which often contains irregular timing, breath-related perturbations, unstable phonation, and disfluency-like events that may affect end-to-end generation and alignment-based recognition differently. Seq2seq models may benefit from more diverse pretraining coverage and stronger decoding priors, whereas SSL+CTC models may be more sensitive under off-the-shelf inference but more responsive to targeted in-domain adaptation. Where available, we include both base- and large-scale variants to examine the effect of model scale in addition to architecture.

\textbf{Evaluation Protocol. }
Our evaluation protocol is designed to characterize post-exercise ASR from three complementary perspectives.
(\textit{i}) \textbf{RQ1: off-the-shelf robustness.} We evaluate off-the-shelf inference on the \texttt{Static} and \texttt{Post-All} settings using all 9 model settings. This provides the main benchmark view of whether post-exercise speech is harder to recognize than \texttt{Static} speech under baseline inference, and how this degradation varies across model families and scales.
(\textit{ii}) \textbf{RQ2: in-domain fine-tuning.} We further fine-tune each model on post-exercise data using 5-fold cross-validation, reporting the mean $\pm$ standard deviation across folds to capture both the average benefit of in-domain supervision and the stability of adaptation.
(\textit{iii}) \textbf{RQ3: impact of fluency.} We conduct an exploratory case study in \texttt{Fluent}/\texttt{Non-Fluent} subgroup analysis to examine whether post-exercise ASR difficulty varies across speakers or interacts with fluency-related factors. Given the small \texttt{Non-Fluent} subset, results are interpreted cautiously and primarily motivate more explicit subgroup-aware evaluation in future work.


\begin{table}[t]
\centering
\caption{ASR models used in this study. Encoder and decoder layers are reported for seq2seq models, while SSL-based CTC models include only encoder layers.}
\label{tab:model_specs}
\small
\setlength{\tabcolsep}{4.5pt}
\renewcommand{\arraystretch}{1.12}

\resizebox{\columnwidth}{!}{
\begin{tabular}{l l c c}
\toprule
\textbf{Size} & \textbf{Model} & \textbf{\# Enc. layers} & \textbf{\# Dec. layers} \\
\midrule
\multicolumn{4}{c}{\textbf{SSL encoder + CTC head}}\\
\midrule
95M   & Wav2Vec2-Base                   & 12 & -- \\
95M   & HuBERT-Base                     & 12 & -- \\
95M   & WavLM-Base+                     & 12 & -- \\
317M  & Wav2Vec2-Large                  & 24 & -- \\
317M  & HuBERT-Large                    & 24 & -- \\
316M  & WavLM-Large                     & 24 & -- \\
\midrule
\multicolumn{4}{c}{\textbf{Transformer seq2seq encoder--decoder}}\\
\midrule
74M   & Whisper-Base            & 6  & 6  \\
1.55B & Whisper-Large-v3      & 32 & 32 \\
\midrule
\multicolumn{4}{c}{\textbf{Paraformer encoder--decoder}}\\
\midrule
220M  & FunASR/Paraformer (L preset) & config-dependent & config-dependent \\
\bottomrule
\end{tabular}
}
\vspace{-0.2in}
\end{table}

\noindent\textbf{Metrics.}
We evaluate recognition quality primarily using Word Error Rate (WER) and Character Error Rate (CER), with lower values indicating better performance. WER is the standard metric for ASR because it reflects word-level substitutions, deletions, and insertions relative to the reference transcript. CER provides a complementary character-level view, which is particularly useful in this setting because post-exercise speech may contain partial words, truncated productions, and local decoding failures that are not always fully reflected at the word level. Using multiple metrics is important in this problem setting because recognition errors on post-exercise speech may arise from several sources, including deletions of weakly articulated words, insertions around breath events, misrecognition of acoustically unstable segments, and distorted handling of repeated or incomplete fragments. Reporting WER and CER, therefore, provides a more complete picture of ASR behavior under post-exercise variation.

\section{Off-the-Shelf Baseline Results}
\label{sec:baseline}
\begin{table*}[t!]
\centering
\caption{Results for all 9 evaluated model settings.Fine-tune uses 5-fold CV on \texttt{Post-All} (mean $\pm$ std). Lower WER/CER is better. Best results in each block are \textbf{bold}; second-best results are \underline{underlined}.}
\label{tab:all9_models_big}
\scriptsize
\renewcommand{\arraystretch}{1.08}
\begin{tabular*}{0.96\textwidth}{@{\extracolsep{\fill}}llcccccc@{}}
\toprule
\textbf{Size} & \textbf{Model}
& \multicolumn{2}{c}{\textbf{\texttt{Static} Baseline}} 
& \multicolumn{2}{c}{\textbf{\texttt{Post-All} Baseline}} 
& \multicolumn{2}{c}{\textbf{\texttt{Post-All} Fine-tune (5-fold)}} \\
\cmidrule(lr){3-4}\cmidrule(lr){5-6}\cmidrule(lr){7-8}
& & \textbf{WER(\%) $\downarrow$} & \textbf{CER(\%) $\downarrow$} & \textbf{WER(\%) $\downarrow$} & \textbf{CER(\%) $\downarrow$}
& \textbf{WER(\%) (mean $\pm$ std) $\downarrow$} & \textbf{CER(\%) (mean $\pm$ std) $\downarrow$} \\
\midrule
74M   & Whisper-Base (S2S)    & 26.86 & 21.48 & 27.48 & 21.87 & $58.11 \pm 8.82$ & $47.62 \pm 5.72$ \\
95M   & Wav2Vec2-Base (CTC)   & 21.53 & 10.11 & 32.30 & 15.70 & $22.93 \pm 1.61$ & $9.31 \pm 0.84$ \\
95M   & HuBERT-Base (CTC)     & 31.74 & 13.68 & 45.98 & 21.04 & $24.75 \pm 0.38$ & $9.33 \pm 0.27$ \\
95M   & WavLM-Base+ (CTC)     & 35.32 & 14.78 & 45.45 & 20.46 & $22.94 \pm 0.43$ & $8.67 \pm 0.29$ \\
\midrule
220M  & FunASR/Paraformer (L preset)     & 12.29 & 6.70  & \textbf{14.57} & \textbf{8.21} & \bm{$13.71 \pm 0.96$} & $8.04 \pm 0.55$ \\
316M  & WavLM-Large (CTC)     & 22.89 & 9.17  & 29.80 & 11.96 & $19.36 \pm 0.28$ & \underline{$6.61 \pm 0.16$} \\
317M  & Wav2Vec2-Large (CTC)  & \textbf{10.57} & \textbf{5.21} & \underline{18.03} & \underline{9.01}  & $32.57 \pm 7.78$ & $13.30 \pm 3.24$ \\
317M  & HuBERT-Large (CTC)    & \underline{11.54} & \underline{5.49} & 19.90 & 9.52  & \underline{$16.53 \pm 0.33$} & \bm{$5.96 \pm 0.24$} \\
1.55B & Whisper-Large-v3 (S2S)& 21.28 & 19.35 & 22.57 & 19.66 & $27.19 \pm 2.53$ & $22.93 \pm 2.52$ \\
\bottomrule
\end{tabular*}
\end{table*}
\noindent\textbf{\texttt{Static} vs.\ \texttt{Post-All} Baseline Comparison. }Following the evaluation protocol described in Section~\ref{sec:experiment}, we first address \textbf{RQ1} from Section~\ref{sec:intro} by examining off-the-shelf baseline robustness on the \texttt{Static} and \texttt{Post-All} settings. Table~\ref{tab:all9_models_big} shows a clear overall pattern: most models perform worse on \texttt{Post-All} speech than on \texttt{Static} speech under baseline inference. The degradation is especially evident for several SSL+CTC models. For example, Wav2Vec2-Base increases from 21.53\% to 32.30\% WER and from 10.11\% to 15.70\% CER, while WavLM-Base+ increases from 35.32\% to 45.45\% WER and from 14.78\% to 20.46\% CER. By contrast, FunASR remains comparatively stable, changing only from 12.29\% to 14.57\% WER and from 6.70\% to 8.21\% CER, and achieves the strongest \texttt{Post-All} baseline result in this benchmark. 
This performance gap mainly stems from how models handle temporal irregularities. Autoregressive models like Whisper tend to smooth or collapse disrupted timing due to strong language priors, while alignment-based approaches such as CTC and Paraformer better preserve duration–token correspondence and remain more robust. Additionally, FunASR benefits from large-scale, real-world training data, further improving its stability under such conditions.

\noindent\textbf{Architecture- and Scale-Level Comparison. }Table~\ref{tab:all9_models_big} also provides an architecture- and scale-level view of baseline performance across all 9 evaluated model settings. FunASR is the strongest off-the-shelf model on \texttt{Post-All}, achieving 14.57\% WER and 8.21\% CER. Among the SSL+CTC models, larger variants are generally stronger than their base counterparts under baseline inference, for example, Wav2Vec2-Large versus Wav2Vec2-Base (18.03\% vs.\ 32.30\% WER; 9.01\% vs.\ 15.70\% CER) and HuBERT-Large versus HuBERT-Base (19.90\% vs.\ 45.98\% WER; 9.52\% vs.\ 21.04\% CER). WavLM-Large also improves over WavLM-Base+ (29.80\% vs.\ 45.45\% WER; 11.96\% vs.\ 20.46\% CER), although both remain weaker than the strongest models in the benchmark. These results suggest that post-exercise robustness depends on both architecture and scale, and that scale alone does not guarantee the strongest baseline performance.

\noindent\textbf{Baseline Differences Across Fluency.}
The subgroup results provide a second perspective on robustness. In the \texttt{Post-All} setting (Table~\ref{tab:post_fluency_split}), \texttt{Non-Fluent} speech is consistently harder to recognize than \texttt{Fluent} speech across all 9 evaluated model settings. For example, FunASR increases from 14.87\% to 20.00\% WER and from 8.27\% to 11.80\% CER, Wav2Vec2-Base from 32.83\% to 49.41\% WER and from 15.72\% to 24.42\% CER, HuBERT-Base from 47.26\% to 70.52\% WER and from 22.12\% to 35.86\% CER, and WavLM-Large from 29.99\% to 43.97\% WER and from 12.11\% to 18.54\% CER. Whisper shows one of the largest subgroup gaps, with Whisper-Large-v3 increasing from 22.07\% to 49.47\% WER. A similar pattern is already visible in the \texttt{Static} condition (Table~\ref{tab:pre_baseline_fluency_split}). Even before exercise, \texttt{Non-Fluent} speech remains harder to recognize than \texttt{Fluent} speech, for example, for FunASR (11.12\% to 15.00\% WER; 6.13\% to 8.01\% CER), Wav2Vec2-Large (9.40\% to 13.28\% WER; 4.64\% to 6.55\% CER), and WavLM-Large (20.76\% to 27.83\% WER; 8.22\% to 11.37\% CER). This indicates that the subgroup effect is not introduced solely by exercise; rather, post-exercise conditions appear to compound an existing fluency-related source of difficulty. Taken together, the baseline results show two consistent patterns: post-exercise speech is harder to recognize than \texttt{Static} speech, and \texttt{Non-Fluent} speech is harder to recognize than \texttt{Fluent} speech in both \texttt{Static} and post-exercise conditions. Without in-domain fine-tuning, FunASR is the strongest off-the-shelf model in this benchmark.


\section{Fine-Tuning Effects and Fluency Case Study}
\label{sec:ft}
\begin{table*}[t!]
\centering
\caption{Fluency split results on \texttt{Post-All} ($n=509$: Fluent=475, Non-Fluent=34). Baseline = inference only; Fine-tune = evaluation of the corresponding 5-fold-trained model on each subgroup split. Lower WER/CER is better. Best results in each block are \textbf{bold}; second-best results are \underline{underlined}.}
\label{tab:post_fluency_split}
\scriptsize
\renewcommand{\arraystretch}{1.08}
\begin{tabular*}{0.94\textwidth}{@{\extracolsep{\fill}}llcccccccc@{}}
\toprule
& & \multicolumn{4}{c}{\textbf{Baseline (\texttt{Post-All})}} & \multicolumn{4}{c}{\textbf{Fine-tune (\texttt{Post-All}, 5-fold)}} \\
\cmidrule(lr){3-6}\cmidrule(lr){7-10}
\textbf{Size} & \textbf{Model}
& \multicolumn{2}{c}{\textbf{\texttt{Fluent}}} & \multicolumn{2}{c}{\textbf{\texttt{Non-Fluent}}}
& \multicolumn{2}{c}{\textbf{\texttt{Fluent}}} & \multicolumn{2}{c}{\textbf{\texttt{Non-Fluent}}} \\
\cmidrule(lr){3-4}\cmidrule(lr){5-6}\cmidrule(lr){7-8}\cmidrule(lr){9-10}
& & \textbf{WER(\%) $\downarrow$} & \textbf{CER(\%) $\downarrow$} & \textbf{WER(\%) $\downarrow$} & \textbf{CER(\%) $\downarrow$}
  & \textbf{WER(\%) $\downarrow$} & \textbf{CER(\%) $\downarrow$} & \textbf{WER(\%) $\downarrow$} & \textbf{CER(\%) $\downarrow$} \\
\midrule
74M   & Whisper-Base       & 27.63 & 20.96 & 49.97 & 48.00 & 65.98 & 54.58 & 95.37 & 82.03 \\
95M   & Wav2Vec2-Base      & 32.83 & 15.72 & 49.41 & 24.42 & 17.33 & 8.71 & 27.99 & 14.38 \\
95M   & HuBERT-Base        & 47.26 & 22.12 & 70.52 & 35.86 & \underline{9.96} & \underline{5.15} & \textbf{16.69} & \textbf{8.64} \\
95M   & WavLM-Base+        & 44.88 & 19.22 & 73.33 & 35.49 & 19.17 & 8.44 & 30.13 & 12.56 \\
\midrule
220M  & FunASR/Paraformer (L preset)  & \textbf{14.87} & \textbf{8.27} & \textbf{20.00} & \textbf{11.80} & 12.78 & 7.56 & \underline{16.84} & 10.21 \\
316M  & WavLM-Large        & 29.99 & 12.11 & 43.97 & 18.54 & 16.77 & 6.70 & 26.21 & 10.56 \\
317M  & Wav2Vec2-Large     & \underline{17.43} & \underline{8.57} & \underline{28.24} & \underline{14.14} & 28.47 & 12.55 & 45.55 & 22.62 \\
317M  & HuBERT-Large       & 19.71 & 9.44 & 33.28 & 17.10 & \textbf{9.55} & \textbf{5.01} & 17.46 & \underline{8.67} \\
1.55B & Whisper-Large-v3   & 22.07 & 18.68 & 49.47 & 46.87 & 32.14 & 26.40 & 52.06 & 43.33 \\
\bottomrule
\end{tabular*}
\end{table*}

\noindent\textbf{Efficacy of In-Domain Fine-Tuning. }
Building on the baseline findings in Section~\ref{sec:baseline}, we next address \textbf{RQ2} by examining whether post-exercise fine-tuning can recover the degradation observed under off-the-shelf inference. Table~\ref{tab:all9_models_big} compares post-exercise baseline performance with 5-fold fine-tuning results for all 9 evaluated model settings. Fine-tuning is often beneficial, but its effect is strongly model-dependent. Several SSL+CTC models improve substantially after adaptation. For example, WavLM-Base+ improves markedly, with WER decreasing from 45.45\% to 22.94\% and CER decreasing from 20.46\% to 8.67\%. Similar gains are observed for Wav2Vec2-Base (32.30\% to 22.93\% WER; 15.70\% to 9.31\% CER), HuBERT-Base (45.98\% to 24.75\% WER; 21.04\% to 9.33\% CER), HuBERT-Large (19.90\% to 16.53\% WER; 9.52\% to 5.96\% CER), and WavLM-Large (29.80\% to 19.36\% WER; 11.96\% to 6.61\% CER). These results suggest that weak off-the-shelf robustness does not necessarily imply weak adaptation potential. FunASR shows a different pattern: its baseline performance is already strong, and fine-tuning yields only modest additional gains, with WER decreasing from 14.57\% to 13.71\% and CER from 8.21\% to 8.04\%. By contrast, adaptation is not uniformly reliable. Whisper-Base performs substantially worse after fine-tuning, with WER increasing from 27.48\% to 58.11\% and CER increasing from 21.87\% to 47.62\%. Whisper-Large-v3 and Wav2Vec2-Large also deteriorate after adaptation. Overall, in-domain adaptation can substantially improve post-exercise ASR, but its effectiveness depends strongly on architecture, model scale, and training stability.

\noindent\textbf{Impact of Fluency on Model Robustness. }
We next address \textbf{RQ3} by examining whether the same pattern holds across \texttt{Fluent} and \texttt{Non-Fluent} speakers. This analysis is exploratory because the post-exercise \texttt{Non-Fluent} subset is relatively small ($n=34$). Under baseline inference, a highly consistent pattern emerges: \texttt{Non-Fluent} speech is harder to recognize than \texttt{Fluent} speech across all 9 model settings. For example, FunASR performs worse on the \texttt{Non-Fluent} subset, with WER/CER increasing from 14.87\%/8.27\% to 20.00\%/11.80\%. Similar subgroup gaps are observed for Wav2Vec2-Base (32.83\%/15.72\% to 49.41\%/24.42\%), HuBERT-Base (47.26\%/22.12\% to 70.52\%/35.86\%), and WavLM-Large (29.99\%/12.11\% to 43.97\%/18.54\%). Whisper exhibits one of the largest subgroup disparities, with Whisper-Large-v3 worsening from 22.07\%/18.68\% to 49.47\%/46.87\%. This consistency suggests that subgroup difficulty is a general property of the benchmark rather than an artifact of a single architecture.

Fine-tuning improves performance for both \texttt{Fluent} and \texttt{Non-Fluent} speakers for several models, although the magnitude and stability of these gains vary across architectures. The strongest improvements are observed for HuBERT and WavLM. HuBERT-Large achieves the best fine-tuned result on the \texttt{Fluent} subset (9.55\% WER, 5.01\% CER), while HuBERT-Base achieves the best fine-tuned result on the \texttt{Non-Fluent} subset (16.69\% WER, 8.64\% CER). WavLM-Large also improves substantially, with WER/CER decreasing from 29.99\%/12.11\% to 16.77\%/6.70\% on \texttt{Fluent} speech and from 43.97\%/18.54\% to 26.21\%/10.56\% on \texttt{Non-Fluent} speech. FunASR remains strong and improves modestly but consistently, with WER/CER decreasing from 14.87\%/8.27\% to 12.78\%/7.56\% on \texttt{Fluent} speech and from 20.00\%/11.80\% to 16.84\%/10.21\% on \texttt{Non-Fluent} speech. At the same time, subgroup-level gains are not universal: Wav2Vec2-Large worsens after fine-tuning on both subsets, and Whisper remains unstable across both groups.

\begin{table}[t!]
\centering
\caption{Baseline fluency split results on \texttt{Static}. This reference set contains 198 structured paragraph-reading recordings (Fluent=161,Non-Fluent=37).}
\label{tab:pre_baseline_fluency_split}
\scriptsize
\setlength{\tabcolsep}{4pt}
\renewcommand{\arraystretch}{1.12}
\resizebox{\linewidth}{!}{
\begin{tabular}{llcccc}
\toprule
& & \multicolumn{2}{c}{\textbf{\texttt{Static Fluent}}} & \multicolumn{2}{c}{\textbf{\texttt{Static Non-Fluent}}} \\
\cmidrule(lr){3-4}\cmidrule(lr){5-6}
\textbf{Size} & \textbf{Model} & \textbf{WER (\%) $\downarrow$} & \textbf{CER (\%) $\downarrow$} & \textbf{WER (\%) $\downarrow$} & \textbf{CER (\%) $\downarrow$} \\
\midrule
74M   & Whisper-Base      & 23.83 & 19.02 & 33.89 & 27.17 \\
95M   & Wav2Vec2-Base     & 19.47 &  9.25 & 26.28 & 12.11 \\
95M   & HuBERT-Base       & 29.20 & 12.46 & 37.85 & 16.64 \\
95M   & WavLM-Base+       & 31.97 & 13.44 & 43.09 & 17.90 \\
\midrule
220M  & FunASR/Paraformer (L preset) & 11.12 &  6.13 & 15.00 &  8.01 \\
316M  & WavLM-Large       & 20.76 &  8.22 & 27.83 & 11.37 \\
317M  & Wav2Vec2-Large    &  \textbf{9.40} &  \textbf{4.64} & \textbf{13.28} &  \textbf{6.55} \\
317M  & HuBERT-Large      & \underline{10.24} &  \underline{4.95} & \underline{14.55} &  \underline{6.74} \\
1.55B & Whisper-Large-v3  & 16.77 & 15.11 & 31.72 & 29.17 \\
\bottomrule
\end{tabular}
}
\end{table}

To interpret the subgroup pattern more carefully, we also examine the pre/static split in Table~\ref{tab:pre_baseline_fluency_split}. Even before exercise, \texttt{Non-Fluent} speech is harder to transcribe than \texttt{Fluent} speech. For example, FunASR increases from 11.12\% to 15.00\% WER and from 6.13\% to 8.01\% CER, Wav2Vec2-Large from 9.40\% to 13.28\% WER and from 4.64\% to 6.55\% CER, and WavLM-Large from 20.76\% to 27.83\% WER and from 8.22\% to 11.37\% CER. This indicates that the \texttt{Fluent}/\texttt{Non-Fluent} gap is not generated solely by exercise; rather, post-exercise conditions appear to compound an existing fluency-related source of difficulty. Taken together, these subgroup results support a conservative interpretation. \texttt{Non-Fluent} speech is consistently harder to recognize than \texttt{Fluent} speech in both pre/static and post-exercise conditions, and several adapted models---especially HuBERT, WavLM, and FunASR---show encouraging gains on the \texttt{Non-Fluent} subset. However, because the post-exercise \texttt{Non-Fluent} subset is small and some model settings remain unstable, these findings should be treated as exploratory rather than definitive. The main implication is methodological: future post-exercise ASR evaluation should explicitly distinguish fluency-related variation from exercise-induced speech variation.
\section{Discussion and Conclusion}~\label{sec:discussion}
Taken together, the findings in Sections~\ref{sec:baseline} and~\ref{sec:ft} support three main conclusions. (\emph{i}) Off-the-shelf ASR is not uniformly robust to post-exercise speech. Across the \texttt{Static}/\texttt{Post-All} benchmark, most models perform worse after exercise, indicating that exercise-related changes in breathing, pausing, and phonation create a practically meaningful distribution shift for transcription. (\emph{ii}) Off-the-shelf robustness and adaptation potential are not the same. FunASR is the strongest and most stable off-the-shelf model in our benchmark, making it a strong default choice when no in-domain adaptation is available. However, several SSL+CTC models, especially HuBERT and WavLM, benefit substantially from post-exercise fine-tuning, showing that baseline strength alone does not determine adaptation behavior. (\emph{iii}) The \texttt{Fluent}/\texttt{Non-Fluent} analysis should be interpreted as an exploratory case study rather than a definitive subgroup claim. \texttt{Non-Fluent} speech is consistently harder to recognize than \texttt{Fluent} speech in both pre/static and post-exercise conditions, and several adapted models show encouraging gains on this subset. At the same time, the post-exercise \texttt{Non-Fluent} subset is small, so the main takeaway is methodological: subgroup-aware evaluation matters for interpreting post-exercise ASR robustness.

\noindent\textbf{Limitations and Future Directions.} This study is limited by the small non-fluent post-exercise subset, the English-only and modest-scale datasets, and the instability of some subgroup-level and fine-tuning results. Future work should therefore collect more balanced and diverse post-exercise speech data, develop more stable adaptation strategies, analyze insertion, deletion, and substitution errors around breaths, pauses, repetitions, and truncated productions in greater detail, and more clearly separate exercise-induced variation from fluency-related variation. In this paper, we benchmarked 9 ASR model settings on post-exercise speech under both off-the-shelf inference and in-domain fine-tuning. Our results show that post-exercise speech remains challenging for current ASR systems, that off-the-shelf robustness is strongly model-dependent, and that fine-tuning can substantially improve some model families while destabilizing others. We further presented a fluent/non-fluent case study showing that non-fluent speech is consistently harder to recognize and should be considered explicitly when interpreting robustness. Overall, our findings provide a practical benchmark for post-exercise ASR and suggest that future work should be both adaptation-aware and subgroup-aware.

\bibliographystyle{IEEEtran}
\bibliography{reference}

@article{zhou2025voice,
  title={Voice as a sensitive biomarker for predicting exercise intensity: a modelling study},
  author={Zhou, Shuyi and Ma, Ruisi and Hu, Wangjing and Zhang, Dandan and Hu, Rui and Zou, Shengwei and Cai, Dingyi and Jiang, Zikang and Ding, Hexiao and Liu, Ting},
  journal={Frontiers in Physiology},
  volume={16},
  pages={1483828},
  year={2025},
  publisher={Frontiers Media SA}
}

@inproceedings{mujtaba2024lost,
  title={Lost in transcription: Identifying and quantifying the accuracy biases of automatic speech recognition systems against disfluent speech},
  author={Mujtaba, Dena and Mahapatra, Nihar and Arney, Megan and Yaruss, J and Gerlach-Houck, Hope and Herring, Caryn and Bin, Jia},
  booktitle={Proceedings of the 2024 Conference of the North American Chapter of the Association for Computational Linguistics: Human Language Technologies (Volume 1: Long Papers)},
  pages={4795--4809},
  year={2024}
}

@inproceedings{nie2025multimodal,
  author    = {Jingping Nie and Yuxuan Fan and Mingyu Zhao and Rui Wan and Ziyi Xuan and Maximilian Preindl and Xiaoyi Jiang},
  title     = {Multi-Modal Dataset Across Exertion Levels: Capturing Post-Exercise Speech, Breathing, and Phonocardiogram},
  booktitle = {Proceedings of the 23rd ACM Conference on Embedded Networked Sensor Systems},
  pages     = {297--304},
  year      = {2025},
  doi       = {10.1145/3715014.3722065}
}

@article{mitra2024breathing,
  author  = {Vikramjit Mitra and Anirban Chatterjee and Ke Zhai and Helen Weng and Ayuko Hill and Nicole Hay and Christopher Webb and Jamie Cheng and Erdrin Azemi},
  title   = {Pre-Trained Foundation Model Representations to Uncover Breathing Patterns in Speech},
  journal = {arXiv preprint arXiv:2407.13035},
  year    = {2024}
}

@article{nie2025soundtrack,
  title={SoundTrack: A Contactless Mobile Solution for Real-time Running Metric Estimation for Treadmill Running in the Wild},
  author={Nie, Jingping and Fan, Yuang and Xuan, Ziyi and Zhao, Minghui and Wan, Runxi and Preindl, Matthias and Jiang, Xiaofan},
  journal={Proceedings of the ACM on Interactive, Mobile, Wearable and Ubiquitous Technologies},
  volume={9},
  number={2},
  pages={1--30},
  year={2025},
  publisher={ACM New York, NY, USA}
}

@inproceedings{lea2021sep28k,
  author    = {Colin Lea and Vikramjit Mitra and Aparna Joshi and Sachin Kajarekar and Jeffrey P. Bigham},
  title     = {{SEP-28k}: A Dataset for Stuttering Event Detection From Podcasts With People Who Stutter},
  booktitle = {ICASSP 2021 - 2021 IEEE International Conference on Acoustics, Speech and Signal Processing (ICASSP)},
  pages     = {6798--6802},
  year      = {2021}
}

@article{ratner2018fluencybank,
  author  = {Nan Bernstein Ratner and Brian MacWhinney},
  title   = {FluencyBank: A New Resource for Fluency Research and Practice},
  journal = {Journal of Fluency Disorders},
  volume  = {56},
  pages   = {69--80},
  year    = {2018}
}

@inproceedings{lea2023enabling,
  author    = {Colin Lea and Zifang Huang and Justin Narain and Lyle Tooley and Derek Yee and Daniel T. Tran and Panayiotis Georgiou and Jeffrey P. Bigham and Leah Findlater},
  title     = {From User Perceptions to Technical Improvement: Enabling People Who Stutter to Better Use Speech Recognition},
  booktitle = {Proceedings of the 2023 CHI Conference on Human Factors in Computing Systems},
  year      = {2023},
  doi       = {10.1145/3544548.3581224}
}

@inproceedings{mujtaba2024inclusive,
  author    = {Dena Mujtaba and Nihar R. Mahapatra and Megan Arney and J. Scott Yaruss and Caryn Herring and Jia Bin},
  title     = {Inclusive {ASR} for Disfluent Speech: Cascaded Large-Scale Self-Supervised Learning with Targeted Fine-Tuning and Data Augmentation},
  booktitle = {Proc. Interspeech 2024},
  pages     = {1275--1279},
  year      = {2024},
  doi       = {10.21437/Interspeech.2024-2246}
}

@inproceedings{mujtaba2025finetuning,
  author    = {Dena Mujtaba and Nihar R. Mahapatra},
  title     = {Fine-Tuning {ASR} for Stuttered Speech: Personalized vs. Generalized Approaches},
  booktitle = {Proc. Interspeech 2025},
  pages     = {3568--3572},
  year      = {2025},
  doi       = {10.21437/Interspeech.2025-2373}
}

@article{radford2022whisper,
  author  = {Alec Radford and Jong Wook Kim and Tao Xu and Greg Brockman and Christine McLeavey and Ilya Sutskever},
  title   = {Robust Speech Recognition via Large-Scale Weak Supervision},
  journal = {arXiv preprint arXiv:2212.04356},
  year    = {2022}
}

@article{gao2022paraformer,
  author  = {Zhifu Gao and Shiliang Zhang and Ian McLoughlin and Zhijie Yan},
  title   = {Paraformer: Fast and Accurate Parallel Transformer for Non-Autoregressive End-to-End Speech Recognition},
  journal = {arXiv preprint arXiv:2206.08317},
  year    = {2022}
}

@article{baevski2020wav2vec2,
  author  = {Alexei Baevski and Henry Zhou and Abdelrahman Mohamed and Michael Auli},
  title   = {wav2vec 2.0: A Framework for Self-Supervised Learning of Speech Representations},
  journal = {arXiv preprint arXiv:2006.11477},
  year    = {2020}
}

@article{hsu2021hubert,
  author  = {Wei-Ning Hsu and Benjamin Bolte and Yao-Hung Hubert Tsai and Kushal Lakhotia and Ruslan Salakhutdinov and Abdelrahman Mohamed},
  title   = {{HuBERT}: Self-Supervised Speech Representation Learning by Masked Prediction of Hidden Units},
  journal = {IEEE/ACM Transactions on Audio, Speech, and Language Processing},
  volume  = {29},
  pages   = {3451--3460},
  year    = {2021}
}

@article{chen2021wavlm,
  author  = {Sanyuan Chen and Chengyi Wang and Zhengyang Chen and Yu Wu and Shujie Liu and Zhuo Chen and Jinyu Li and Naoyuki Kanda and Takuya Yoshioka and Xiong Xiao and Jian Wu and Long Zhou and Shuo Ren and Yanmin Qian and Yao Qian and Jian Wu and Michael Zeng and Xiangzhan Yu and Furu Wei},
  title   = {{WavLM}: Large-Scale Self-Supervised Pre-Training for Full Stack Speech Processing},
  journal = {arXiv preprint arXiv:2110.13900},
  year    = {2021}
}

@inproceedings{wang2025breathing,
  title={Breathing and Semantic Pause Detection and Exertion-Level Classification in Post-Exercise Speech},
  author={Wang, Yuyu and Xia, Wuyue and Yao, Huaxiu and Nie, Jingping},
  booktitle={Proceedings of the 3rd ACM International Workshop on Intelligent Acoustic Systems and Applications},
  pages={13--18},
  year={2025}
}

@article{milne2020effectiveness,
  title={The effectiveness of artificial intelligence conversational agents in health care: systematic review},
  author={Milne-Ives, Madison and De Cock, Caroline and Lim, Ernest and Shehadeh, Melissa Harper and De Pennington, Nick and Mole, Guy and Normando, Eduardo and Meinert, Edward},
  journal={Journal of medical Internet research},
  volume={22},
  number={10},
  pages={e20346},
  year={2020},
  publisher={JMIR Publications Inc., Toronto, Canada}
}

@article{xue2026edge,
  title={Edge-Cloud Collaborative Speech Emotion Captioning via Token-Level Speculative Decoding in Audio-Language Models},
  author={Xue, Xiangyuan and Lu, Jiajun and Gao, Yan and Huang, Gongping and Dang, Ting and Jia, Hong},
  journal={arXiv preprint arXiv:2603.11397},
  year={2026}
}

@article{xia2025convergence,
  title={The Convergence of Mental Health and AI: A Cross-Disciplinary Survey of Ubiquitous Sensing, LLMs, and Clinical Alignment},
  author={Xia, Wuyue and Shao, Hanya and Kong, Ningxin and Fan, Yuang and Nie, Jingping},
  journal={Authorea Preprints},
  year={2025},
  publisher={Authorea}
}

@article{nie2025foundation,
  title={Foundation Model Hidden Representations for Heart Rate Estimation from Auscultation},
  author={Nie, Jingping and Tran, Dung T and Thakkar, Karan and Kowtha, Vasudha and Huang, Jon and Avendano, Carlos and Azemi, Erdrin and Mitra, Vikramjit},
  journal={Interspeech 2025},
  year={2025}
}

@inproceedings{mitra2024investigating,
  author={Mitra, Vikramjit and Nie, Jingping and Azemi, Erdrin},
  booktitle={Proc. IEEE International Conference on Acoustics, Speech and Signal Processing (ICASSP'24)}, 
  title={Investigating Salient Representations and Label Variance in Dimensional Speech Emotion Analysis}, 
  year={2024}
}

@article{mitra2025leveraging,
  title={Leveraging saliency-based pre-trained foundation model representations to uncover breathing patterns in speech},
  author={Mitra, Vikramjit and Chatterjee, Anirban and Zhai, Ke and Weng, Helen and Hill, Ayuko and Hay, Nicole and Webb, Christopher and Cheng, Jamie and Azemi, Erdrin},
  journal={Computer Speech \& Language},
  pages={101926},
  year={2025},
  publisher={Elsevier}
}

\end{document}